\documentstyle[12pt]{article}

\def\ga{\alpha}
\def\bs{{\bf s}}
\def\half{\frac{1}{2}}
\newcommand{\be}{\begin{equation}}
\newcommand{\ee}{\end{equation}}

\newcommand{\bea}{\begin{eqnarray}}
\newcommand{\bee}{\begin{eqnarray}}
\newcommand{\eee}{\end{eqnarray}}
\newcommand{\eea}{\end{eqnarray}}
\def\hn{\hat{n}}
\def\hm{\hat{m}}
\def\hk{\hat{k}}
\def\hl{\hat{l}}

\def\ATMP#1#2#3{{\it Adv. Theor. Math. Phys.} {\bf #1} {(#2)} {#3}}

\def\NPB#1#2#3{{\it Nucl. Phys.} {\bf B#1} {(#2)} {#3}}
\def\PLB#1#2#3{{\it Phys. Lett.} {\bf B#1} {(#2)} {#3}}
\def\CQG#1#2#3{{\it Class. Quantum Grav.} {\bf #1} {(#2)} {#3}}
\def\MPLA#1#2#3{{\it Mod. Phys. Lett.} {\bf A#1} {(#2)} {#3}}

\def\PRD#1#2#3{{\it Phys. Rev.} {\bf D#1} {(#2)} {#3}}
\def\PR#1#2#3{{\it Phys. Rep.} {\bf #1} {(#2)} {#3}}
\def\AP#1#2#3{{\it Ann. Phys.} {\bf #1} {(#2)} {#3}}
\def\JMP#1#2#3{{\it J. Math. Phys.} {\bf #1} {(#2)} {#3}}
\def\CMP#1#2#3{{\it Comm. Math. Phys.} {\bf #1} {(#2)} {#3}}

\renewcommand{\thefootnote}{\fnsymbol{footnote}}
\newcommand{\newsection}{\setcounter{equation}{0}\section}
\def\appendix#1{\addtocounter{section}{1}
\setcounter{equation}{0}\renewcommand{\thesection}{\Alph{section}}
\section*{Appendix \thesection\protect\indent \parbox[t]{11.15cm}{#1}}
\addcontentsline{toc}{section}{Appendix \thesection\ \ \ #1}}

\begin{document}

{}~\hfill G\"oteborg ITP

{}~\hfill FIAN/TD-2000-17

{}~\hfill OHSTPY-HEP-T-00-007

{}~\hfill AEI-2000-025

{}~\hfill hep-th/0005136

\vspace{1cm}

\begin{center}
{\large\bf
HOW MASSLESS ARE MASSLESS FIELDS IN $AdS_d$}

\vspace{1.5cm}

L. Brink${}^{{\rm a}}$,
R.R. Metsaev${}^{{\rm b},{\rm c}}$
and M.A. Vasiliev${}^{{\rm d}}$
\footnote{On leave of absence from Lebedev Physical Institute, Moscow
}

\vspace{1cm}
${}^{{\rm a}}${\it
Department of Theoretical Physics,
Chalmers University of Technology, S-412 96, G\"oteborg, Sweden}

\vspace{0.5cm}
${}^{{\rm b}}${\it Department of Theoretical Physics, P.N. Lebedev
Physical Institute, Leninsky prospect 53, 117924, Moscow, Russia}

\vspace{0.5cm}
${}^{{\rm c}}${\it Department of Physics, The Ohio State
University, Columbus, OH 43210-1106, USA}

\vspace{0.5cm}
${}^{{\rm d}}$
{\it Albert-Einstein-Institut,  Max-Planck-Institut f\"ur
Gravitationsphysik,
Am M\"uhlenberg 1, D-14476 Golm, Germany}

\vspace{1.5cm}
{\bf Abstract}
\end{center}

\bigskip

Massless fields of generic Young symmetry type
in $AdS_d$ space are analyzed. It is demonstrated that in contrast to
massless fields in Minkowski space whose physical degrees of freedom
transform in irreps of $o(d-2)$ algebra, $AdS$ massless mixed symmetry
fields reduce to a number of irreps of $o(d-2)$ algebra. From the
field theory perspective this means that
not every massless field in flat space admits
a deformation to $AdS_d$ with the same number of degrees of
freedom, because it is impossible to keep all of the flat space
gauge symmetries unbroken in the AdS space. An equivalent statement
is that, generic irreducible AdS massless fields reduce to certain
reducible sets
of massless fields in the flat limit. A  conjecture on the general
pattern of the flat space limit of a general $AdS_d$ massless field is
made.
The example of the three-cell ``hook'' Young diagram is discussed
in detail. In particular, it is shown that only a combination of the
three-cell flat-space field with a graviton-like field admits a smooth
deformation to $AdS_d$.


\renewcommand{\thefootnote}{\arabic{footnote}}
\setcounter{footnote}{0}

\newpage

\newsection{Introduction}

The aim of this paper is to demonstrate some peculiarities
of generic free massless fields in anti-de Sitter (AdS) space
of an arbitrary space-time dimension $d$. The main conclusion will
be that, in general, an irreducible $AdS_d$ massless field does
not classify according to irreducible representations of the flat
space massless little algebra $o(d-2)$, but reduces to a certain
set of irreducible flat space massless fields. The pattern of necessary
flat-space massless fields will be given.
Another (rather unexpected)  manifestation of this fact
is that not every massless field in flat space admits
a deformation to $AdS_d$ with the same number of degrees of
freedom, since it is impossible to keep all the flat space
gauge symmetries unbroken in the $AdS_d$ space. This phenomenon does not
take place, though, for all those types of massless fields that appear in the
usual low-energy massless sectors of the superstring models and
supergravities, because it holds only for the representations
of the space-time symmetries  described by non-rectangular
Young diagrams. For the same reason it cannot be observed in $AdS_4$
higher spin gauge theories \cite{Fr,vas} (all massless fields
in $AdS_4$ are described by one-row Young diagrams). The effect
discussed in this paper takes place for  $d\geq 6$.

In superstring theory all types of representations
appear at the higher massive levels. The study of higher spin
gauge theory has two main motivations (see e.g. \cite{vas,met08}):
Firstly to overcome the well-known barrier of $N\leq 8$ in
$d=4$ supergravity models
and, secondly, to investigate if there is a most symmetric phase of
superstring theory that leads to the usual string theory as a
result of a certain spontaneous breakdown
of higher spin gauge symmetries. These two motivations lead in fact
into the same direction because, as shown for the $d=4$ case
\cite{FV,van} higher spin gauge theories require infinite collections
of higher spin gauge fields with infinitely increasing spins.
Another important feature discovered in
\cite{FV1} is that gauge invariant higher spin interactions
require the cosmological constant $\lambda^2$
of the background AdS space to be non-zero to compensate
the  extra length dimensions carried by the higher derivative
interactions required by the higher spin gauge symmetries. (In this
perspective $\lambda$ plays the r\^ole analogous to $\alpha^\prime$ in
superstring theory). The fact that higher spin theories require an AdS
background was regarded as rather surprising until
it was realized  that it plays a distinguished r\^ole
in the superstring theory as well \cite{mal}.

To investigate a possible relationship between the superstring
theory and higher spin theories one has  to build
the higher spin gauge theory in higher dimensions, $d>4$ (e.g.
$d=10,11, ...$).
A conjecture on the possible form of the higher spin
symmetries and equations
of motion for higher spin spin gauge fields was made in
\cite{fer,vas2} as a certain
generalization of the $d=4$ results \cite{op,vas1}
which were proved to describe
interactions of all $d=4$ massless fields.

A generalization like this to
higher dimensions is not straightforward because of the use
of certain auxiliary twistor type variables.
As a starting point, it is therefore important to analyze
more carefully the notion of a general massless field in $AdS_d$.
This is the main goal of this paper.

Another motivation comes from the flat space analysis of certain massless
(nonsupersymmetric) triplets in $d=11$ in \cite{pr},\cite{br}, the dimension
of M-theory, where it
was  shown that there exists an  infinite collection of triplets of higher spin fields having equal
numbers of bosonic and fermionic degrees of freedom. These triplets show
some remarkable properties.
For the first four Dynkin indices the bosonic and the fermionic numbers
match up. This phenomenon
is rather special for 11 dimensions and follows from the fact that the
little group $SO(9)$ is an
equal-rank subgroup of $F_4$, bringing in exceptional groups
into the picture. If these
triplets have anything to
do with higher spin gauge field theories and/or M-theory it is an
interesting  question whether it
is possible to extend the analysis of \cite{br} to the AdS case. That
question in fact triggered this investigation.

\section{Massless Unitary  Representations in $AdS_d$}

{}$AdS_d$ is a $d$-dimensional space-time
with signature $(d-1,1)$ and the group of motions
$SO(d-1,2)$. It is most useful to identify $AdS_d$ with the
universal covering space of the appropriate hyperboloid embedded into
Minkowski space-time with
signature\\
$(d-1,2)$. Physically meaningful relativistic fields in $AdS_d$
are classified
according to the lowest weight unitary representations of
$o(d-1,2)$. Unitarity
implies compatibility with quantum mechanics, while lowest weight
of a unitary representation guarantees that the energy is bounded
from below.

\subsection{General Facts}

The commutation relations of $o(d-1,2)$ are
\be
\label{adscom}
[M_{\hm \hn} ,M_{\hk\hl} ]=i(\eta_{\hn\hk}M_{\hm\hl} -\eta_{\hm\hk}M_{\hn\hl}
-\eta_{\hn\hl}M_{\hm\hk} +\eta_{\hm\hl}M_{\hn\hk} )\,,
\ee
where $\eta_{\hn\hm}=(-,+,\ldots +,- )$ is the flat metric in
the $(d-1,2)$  space,
$\hm ,\hn ,\hk ,\hl = 0 \div d$. The generators $M_{\hm\hn}$
are Hermitian.
Let us  choose the following basis in the algebra:
\be
t_\pm^a = \half (M_0{}^a \pm i M_d{}^a )\,,
\ee
\be
E=M_{0d}\,,\qquad L^{ab}=-iM^{ab}\,,
\ee
where $a,b =1\div d-1$. The commutation relations (\ref{adscom})
take the form
\be
\label{grad}
[E ,t_\pm^a ]= \pm t_\pm^a \,,
\ee
\be
\label{tt}
[t_-^a , t_+^b ] = \half (E \delta ^{ab} - L^{ab} )\,,
\ee
\be
[L_{ab} , t_{\pm c } ] = \delta _{bc} t_{\pm a}- \delta _{ac} t_{\pm b}\,,
\ee
\be
[L_{ab},L_{ ce } ]= \delta _{bc} L_{ ae} -\delta _{ac} L_{ be}
- \delta _{be} L_{ ac} +\delta _{ae} L_{ bc}
\ee
with all other commutators vanishing. The hermiticity conditions are
\be
E^\dagger = E\,,\qquad (t_\pm^a)^\dagger = t_\mp^a \,,\qquad
(L^{ab})^\dagger = -L^{ab}\,.
\ee

The generators $E$ and $L^{ab}$ can be identified with the energy and
angular momenta,
respectively,
and span the Lie algebra $o(2)\oplus o(d-1)$
of the maximal compact subgroup of the AdS group $SO(2,d-1)$.
The non-compact generators $t_\pm ^a$ are combinations of $AdS$
translations and Lorentz boosts. The commutation relations are
explicitly $Z$-graded with $t^a_\pm$ having grade $\pm 1$.
$E$ is the grading operator. The lowest weight unitary
representations are now constructed in the standard fashion
(for a review of the $AdS_4$ case see e.g. \cite{Nic} and \cite{dewit})
starting
with the vacuum space $|E_0 ,\bs\rangle$ that is annihilated by
$t_-^a$
\be
\label{t-vac}
t_-^a |E_0 ,\bs\rangle =0
\ee
and forms a unitary representation of the compact subalgebra
$o(2)\oplus o(d-1)$ that means in particular that
\be
\label{E0vac}
E |E_0 ,s\rangle =E_0 |E_0 ,\bs\rangle\,.
\ee
Here $\bs$ denotes the type of representation of $o(d-1)$
carried by the vacuum space:
${\bf s}=(s_1,\ldots s_\nu)$ with $\nu=[\frac{d-1}{2}]$.
${\bf s}$ is a generalized spin characterizing the representation.
In terms of Young tableaux  $s_i$ is the
number of cells in the $i$-th row of the Young tableaux.
Since we are talking about representations of orthogonal algebras,
the corresponding tensors are traceless. We will here not
discuss the self-dual representations that can be
singled out  with the aid of the Levi-Civita symbol\footnote{The
self-dual and antiself-dual representations which appear for odd $d$
are usually distinguished by a sign of the $s_{\frac{d-1}{2}}$.}.
Note that ${\bf s}$ describes a finite-dimensional representation
of $o(d-1)$.  Field-theoretically this corresponds to a finite-component
field carrying a finite spin.

The full representation of the Lie algebra
$o(d-1,2)$, denoted  in \cite{F1} $D(E_0, {\bf s})$,
is spanned by the vectors of the form
\be
\label{mod}
t_+^{a_1} \ldots t_+^{a_k} |E_0 ,\bs\rangle
\ee
for all $k$.
The states with fixed $k$ are called level-$k$ states.
Note that states with pairwise different $k$ are orthogonal as a
consequence
of (\ref{grad}). As a result, the analysis of unitarity can be
reformulated as the check of the positivity of the norms of
the finite-dimensional subspaces at every level.
For sufficiently large generic $E_0$ it is intuitively clear that the
representation $D(E_0, {\bs})$
is irreducible and unitary,  i.e. all norms are strictly positive.
 Such representations are identified with massive representations of
$AdS_d$.

Let us emphasize that the elements of the module (\ref{mod})
can be identified with the modes of a one-particle state in the
corresponding free quantum field theory. The elements of the
$AdS_d$ algebra $o(d-1,2)$ are then realized as bilinears in the
quantum fields.
Note that at the level of equations of motion
the light-cone field-theoretical realization of generic
$AdS_d$ massive representations for arbitrary $E_0$ and ${\bf s}$
has been developed recently in \cite{metlc}.

We see that massive states are classified by the parameter
$E_0$ (which is the analog
of mass) and a representation of $o(d-1)$.
This picture is in
agreement with the standard description of massive relativistic fields
in flat space-time in terms of the Wigner little group $SO(d-1)$.

If $E_0$ gets sufficiently small, the norm cannot stay positive as is
most obvious from the following consequence of (\ref{tt})
\be
[t_-{}_a , t_+^a ] = \frac{d-1}{2} E \,,
\ee
which implies that for negative $E_0$ some
level 1 states cannot have positive norm  for a positive-definite
 vacuum subspace.  There is
therefore a boundary of the unitarity
region $E_0 = E_0 ({\bs}) >0$ such that some states acquire negative
norm for $E_0 < E_0 ({\bs})$. Obviously, these states should
have zero norm for $E_0 = E_0 ({\bs})$.

Starting from the inside of the unitarity region
decreasing $E_0$ for some fixed $\bs$
one approaches the boundary of the
unitarity region, $E_0 = E_0 ({\bs})$.
Some zero-norm vectors then appear for $E_0 = E_0 ({\bs})$.
These necessarily should have
vanishing scalar product to any other state.
(Otherwise, one can
build a negative norm state that is in contradiction with the
assumption that we are at the boundary of the unitarity region).
Therefore, the zero-norm states form an invariant subspace called a
singular submodule. By factoring out this subspace one is left with
a unitary representation which is ``shorter'' than the generic massive
representation.

The resulting ``shortened'' unitary representations
correspond either to massless fields \cite{evans} or to singletons
\cite{dirac} and doubletons \cite{murat},\cite{gunmin}
identified with the conformal
fields at the boundary of the $AdS$ space \cite{dirac}. The fact
that a singular submodule can be factored out admits an interpretation as some
sort of a gauge symmetry (true gauge symmetry for the case of massless
fields, or independence of bulk degrees of freedom for singletons and
doubletons). For this reason we choose this definition of masslessness
for all fields in
$AdS_d$ except for the scalar and spinor massless matter fields which are
not associated with any gauge symmetry principle and
singletons\footnote{Singleton-type fields  live at the boundary of
$AdS_d$ and cannot be interpreted as bulk massless fields.
Presumably, all singletons except for scalar and spinor
correspond to maximally antisymmetrized representations of the $AdS_d$
algebra equivalent to their duals. In particular, this is true for
the second rank antisymmetric tensor representation in the $AdS_5$
case that can be identified with the field strength of the Yang-Mills
fields of the $N=4$ $SYM$ at the boundary of $AdS_5$.}.

The analysis of positive definiteness of the scalar product
of level-1 states was done in \cite{met95} for an arbitrary
even $d$ and an arbitrary type
of representation of  $\bs$ carried by the vacuum state.
The final result is
\be
\label{ub}
E_0 \geq E_0 ({\bf s})\,,
\ee
\be
\label{e0}
E_0 ( {\bf s})= s_1+d-t_1-2\,,
\ee
where $t_1$ is the number of rows of the maximal length $s_1$, i.e.
$$
s_1=\ldots= s_{t_1 -1}=s_{t_1}> s_{t_1 +1} \geq s_{t_1 +2}\geq  \ldots
\geq s_\nu \,.
$$
 It can be shown that the same result is true for odd $d$
(provided that $s_\nu$ is replaced by $|s_\nu|$).
Note that this bound for $d=4$ was originally
found in \cite{evans}. For the case $d=5$ see \cite{mack} and
references therein.

Massless representations are ``shorter'' than massive ones
classified  according to  the parameter $E_0$ (equivalent
of mass) and a representation of $o(d-1)$.
In flat space, massless fields are classified according to the representation
of the massless little group $SO(d-2)$. The question we address here
is whether or not the shortening in $AdS_d$  can be
interpreted in terms of irreducible representations of $SO(d-2)$.
We will show that the answer is no for a generic representation.
This will be demonstrated both at the algebraic level
using the language of singular vectors and at the field-theoretical
level focusing on the simplest nontrivial massless
field with $\bs = (2,1,0,\ldots, 0)$.
The most important field-theoretical conclusion is that
a generic irreducible massless field in $AdS_d$ decomposes into a
collection of massless fields in the flat limit. In that sense, a massless
field in the $AdS_d$ is generically
``less massless'' than elementary massless field in flat space. An
important consequence of this
fact is that not every massless field in flat space can be deformed into
AdS geometry.

\subsection{Singular Vectors}
\label{Singular Vectors}

It is useful to reformulate
the problem in terms of singular vectors. Since the energy $E$ is
bounded from below for the whole representation,
the singular submodule
spanned by zero-norm states is itself a lowest weight representation.
Therefore it contains at least one nontrivial subspace
$|E^\prime_0 ,\bs^\prime \rangle $
that has the properties analogous to (\ref{t-vac}) and (\ref{E0vac}),
\be
\label{sing}
t_-^a  |E^\prime_0 ,\bs^\prime \rangle =0
\ee
and
\be
E |E^\prime_0 ,\bs^\prime \rangle=  E^\prime_0
|E^\prime_0 ,\bs^\prime \rangle\,,
\ee
i.e. it forms some irreducible representation $\bs^\prime$
of $o(d-1)$.
Obviously,
\be
E^\prime_0 = E_0 + k
\ee
if $|E^\prime_0 ,\bs^\prime \rangle$ belongs to the level-$k$ subspace.

Such spaces $|E^\prime_0 ,\bs^\prime \rangle$ we will call singular
vacuum spaces while any of their elements will be called a
singular vector\footnote{This terminology is very closely
although not exactly coinciding with that used for the Verma module
construction when irreducible vacuum subspaces are one-dimensional
because the grade zero subalgebra, namely the Cartan subalgebra, is Abelian.}.
Clearly, singular vacuum spaces
form representations of the algebra $o(2)\oplus o(d-1)$
and therefore decompose into a direct sum of irreducible representations
of $o(d-1)$ on different levels.
The standard situation is with a single irreducible singular
vacuum space.
The singular module as a whole then has the structure
\be
t_+^{a_1} \ldots t_+^{a_k} |E^\prime_0 ,\bs^\prime\rangle\,.
\ee
The factorization to a unitary irreducible representation
is equivalent to identifying all these vectors to zero.

Let us now consider the example of a vacuum state
$|E_0 ,\bs\rangle$ with $\bs=(s_1,s_2,0,\ldots, 0)$
corresponding to a two-row Young diagram,
\bigskip
\be
\label{dia}
\begin{picture}(130,45)
\put(30,34){$s_1$}
{\linethickness{.500mm}
\put(00,20){\line(1,0){140}}%
\put(00,30){\line(1,0){140}}%
\put(00,10){\line(1,0){70}}%
\put(00,10){\line(0,1){20}}%
\put(140,20){\line(0,1){10}}%
\put(70,10){\line(0,1){10}}}%
\put(10,10.0){\line(0,1){20}} \put(20,10.0){\line(0,1){20}}
\put(30,10.0){\line(0,1){20}} \put(40,10.0){\line(0,1){20}}
\put(50,10.0){\line(0,1){20}} \put(60,10.0){\line(0,1){20}}
\put(70,20.0){\line(0,1){10}} \put(80,20.0){\line(0,1){10}}
\put(90,20.0){\line(0,1){10}} \put(100,20.0){\line(0,1){10}}
\put(110,20.0){\line(0,1){10}} \put(120,20.0){\line(0,1){10}}
\put(130,20.0){\line(0,1){10}} 
\put(30,04){$s_2$}
\end{picture}
\ee
\bigskip
with $0\leq s_2\leq s_1 $.

This means that
$|E_0 ,\bs\rangle$  can be realized as a tensor
$v_{a_1 \ldots a_{s_2} ,b_1 \ldots b_{s_1}} (E_0 )$ which is symmetric
both in the
indices $a$ and in the indices $b$, satisfying the antisymmetry property
\be
\label{asym}
v_{  a_2 \ldots a_{s_2} \{b_{s_1 +1}\,,  b_1 \ldots b_{s_1}\}}(E_0 ) =0\,,
\ee
which implies that symmetrization over any $s_1 +1$ indices $a$
and/or $b$ gives zero. The tensor is traceless, which means that
contraction of any two
indices with the $o(d-1)$ invariant flat metric $\eta_{ab}$ gives zero.
Taking (\ref{asym}) into account it is enough to require
\be
\eta^{b_1 b_2} v_{a_1 \ldots a_{s_2} ,b_1 \ldots b_{s_1}} (E_0 ) =0\,.
\ee

The level one states are
\be
\label{cab}
t_+^c v_{a_1 \ldots a_{s_2} ,b_1 \ldots b_{s_1}} (E_0 )\,.
\ee
These states form a reducible representation of $o(d-1)$
(the tensor product
of the vector representation with the representation (\ref{dia})). For
generic $d$, $s_2$ and $s_1$ it contains five irreducible components: two
Young diagrams with one cell less (index $c$ is contracted to either
one of the indices $a$ or one of the indices $b$) and
three diagrams with  one cell more: adding one cell to the first, second
or an (additional) third row. Our problem therefore is to check whether
any of these irreducible representations can be a singular
vacuum space, i.e.
\be
\label{sinv}
t_-^e \Pi_\ga (t_+^c v_{a_1 \ldots a_{s_2} ,b_1 \ldots b_{s_1}}
(E_0 ))\equiv 0
\ee
for some $E_0$ ($\Pi_\ga$ is a  projector to one or another
irreducible component in (\ref{cab})). Let us consider the two
representations with cells cut.
The appropriate projections are given by the
following formulae describing irreducible $o(d-1)$ tensors
\be
v^1_{a_1 \ldots a_{s_2} ,b_1 \ldots b_{s_1 -1}}
= t_+^c \{v_{a_1 \ldots a_{s_2} ,b_1 \ldots b_{s_1 -1}c} (E_0 )
+\frac{s_2}{s_1 -s_2+1}
v_{c a_1 \ldots a_{s_2 -1} ,b_1 \ldots b_{s_1 -1}a_{s_2 }} (E_0 )\}
\ee
(symmetrizations within each of
the groups of indices $a$ and $b$ are assumed)
 and
\be
v^2 _{a_1 \ldots a_{s_2 -1} ,b_1 \ldots b_{s_1}}
= t_+^c v_{c a_1 \ldots a_{s_2 -1} ,b_1 \ldots b_{s_1}} (E_0 ) \,.
\ee
Elementary  computations give that
\begin{eqnarray}
t_{-c}v^1 _{a_1 \ldots a_{s_2} ,b_1 \ldots b_{s_1 -1}}
&=& \half (E_0-(d+s_1 -3)) \Big (
v _{a_1 \ldots a_{s_2} ,b_1 \ldots b_{s_1 -1} c} (E_0 )
\nonumber\\
&+&\frac{s_2}{s_1 -s_2 +1}
v _{a_1 \ldots a_{s_2-1}c ,b_1 \ldots b_{s_1-1} a_{s_2}} (E_0 )\Big )
\end{eqnarray}
and
\be
t_{-c}v^2 _{a_1 \ldots a_{s_2 -1} ,b_1 \ldots b_{s_1}}
=\half (E_0 -(d+s_2 -4))v _{a_1 \ldots a_{s_2 -1}c ,b_1
\ldots b_{s_1}} (E_0 )\,.
\ee
As a result, singular vectors appear at
\be
\label{es}
E_0^1 = d+s_1-3
\ee
and
\be
\label{et}
E_0^2 = d+s_2-4\,.
\ee

A few comments are now in order.

The cases $s_1=s_2$ and $s_1>s_2$ are different because
$v^1 \equiv 0$ for $s_1=s_2$ as a consequence of the antisymmetry property
(\ref{asym}). For $ s_1>s_2$ both $v^1$ and $v^2$ are nontrivial.

As expected, the values of the ``singular'' energies (\ref{es}) and (\ref{et})
are in agreement with the general analysis of \cite{met95},
where it was also shown that only
the representation resulting from the factorization of the singular
submodule with the highest energy of a singular vector
is unitary (this fact is natural from the singular vector description:
unitarity can be preserved only when the boundary of the
unitarity region is approached; this implies the highest $E_0$).
 We therefore conclude that unitary massless particles
appear for
\be
E_0 = d+s_1-3\qquad for\quad s_1>s_2
\ee
and
\be
E_0 = d+s_1-4\qquad for\quad s_1=s_2 \,.
\ee

The analysis in terms of singular vectors
 is simple enough but can be simplified
further with the aid of the technique proposed in \cite{LV}
with the tensors corresponding to various Young diagrams
realized as certain subspaces of an appropriate Fock space.
This technique is explained in section
\ref{Flat Space} since we
will use it in the field-theoretical part of the paper.
We here use the tensor language to make most clear the
interpretation in terms of the representations of
the massless little algebra $o(d-2)$. One can analogously
investigate  singular spaces with the boxes added to
make sure that they have negative energies and therefore
do not play a r\^ole in our analysis.

We expect that the analysis of higher levels does not affect
our conclusions. One reason is that the appearance of singular
vectors at higher levels within the unitarity region
would imply existence of higher spin gauge fields with gauge
transformations having more than one derivative acting on
a gauge parameter. The general analysis of massless fields in
flat space-time of an arbitrary dimension \cite{labas}
shows that this does not take place.

\section{Flat Space Pattern of $AdS$ Massless fields}
\label{Flat Space Pattern of $AdS$ Massless fields}

Let $A_{a_1 \ldots a_{s_1},b_1 \ldots b_{s_2},\ldots }$
be an irreducible tensor of $o(d-1)$ ($a,b\ldots =1\div d-1$)
of a specific symmetry type. Let
$n^a$ be a nonzero vector of $o(d-1)$.   $o(d-2)$ can then
be identified with the stability subalgebra of
$o(d-1)$ that leaves $n^a$ invariant. If the tensor
$A_{a_1 \ldots a_{s_1},b_1 \ldots b_{s_2},\ldots }$
is orthogonal to $n^a$ with respect to all possible
contractions of indices
\be
n^{a_1}A_{a_1 \ldots a_{s_1},b_1 \ldots b_{s_2},\ldots }=0 \,,\qquad
n^{b_1}A_{a_1 \ldots a_{s_1},b_1 \ldots b_{s_2},\ldots }=0 \ldots
\ee
then it describes a representation of $o(d-2)$ of the same symmetry
pattern. If some contractions with $n^a$ are nonzero, one
can decompose the $o(d-1)$ tensor
$A_{a_1 \ldots a_{s_1},b_1 \ldots b_{s_2},\ldots }$ into
irreducible representations of $o(d-2)$ with the aid of the
projection operators constructed from $n^a$ or,
 in other words, performing dimensional reduction. For example,
for a vector,
\be
A^a = A^{\| a} + A^{\bot a}\,,\qquad A^{\bot a} =
A^a -\frac{n^a n_b}{n_c n^c}A^b\,,\qquad
A^{\| a} = \frac{n^a n_b}{n_c n^c}A^b\,.
\ee

The analysis of singular vectors in section
\ref{Singular Vectors} admits a similar interpretation.
Indeed, let us interpret $t_+^a$ as a vector $n^a$ analogous to
the momentum operator in flat-space field theory (note
that the operators $t_+^a$ commute with themselves). The fact
that a singular vector appears means that some contractions of
$t_+^a$ with the vacuum vector decouple from the spectrum and
therefore are equivalent to zero. If all possible contractions
of $t_+^a$ would decouple this would mean that the $o(d-1)$
vacuum representation would reduce to the $o(d-2)$ of the same
symmetry. Since the energies (\ref{es}) and (\ref{et}) are different
this cannot be true simultaneously. Therefore, when a singular vector is
present, the ``reduced representation'' is effectively smaller than
an irreducible representation of    $o(d-1)$ but may be larger than the
corresponding irreducible representation of $o(d-2)$, containing a
number of irreducible representations of $o(d-2)$.

Let us note that the energy in $AdS_d$ is measured in units of the
inverse
$AdS$ radius $\lambda$ that was set equal to unity in
our analysis. Reintroducing $\lambda$ and taking the
flat limit $\lambda \to 0$, all energies of singular vectors tend to zero.
This means that in the flat limit different singular vectors
may decouple simultaneously and therefore a natural possibility
consists of the flat space reduction to one
(totally $t_+^a$ orthogonal) representation of $o(d-2)$,
 in agreement with the standard analysis
\cite{barut} of massless representations of the Poincare' algebra.
However such massless representations of the Poincare' algebra
may not admit a deformation to a representation of the
AdS algebra with $\lambda \neq 0$. At the field-theoretical level
this means that it will not be possible to preserve all necessary
gauge symmetries for $\lambda \neq 0$.
This phenomenon is demonstrated in the field theoretical example in
section \ref{Gauge Symmetries in $AdS_d$}. The deformation will
be possible however, if one starts
with an appropriate collection of massless fields in
flat space dictated by the ``incomplete'' dimensional reduction
via decoupling of singular vectors. The main aim of this section is
to formulate  a conjecture on the pattern of massless fields
in flat space compatible with the deformation to $AdS_d$.

Let us consider an arbitrary Young diagram with
row lengths $s_1 \geq s_2 \geq s_3 \ldots$. For our analysis
it is more convenient to build  Young diagrams not from rows
as elementary entities but from rectangular blocks of
an arbitrary  height $t$ and length $s$

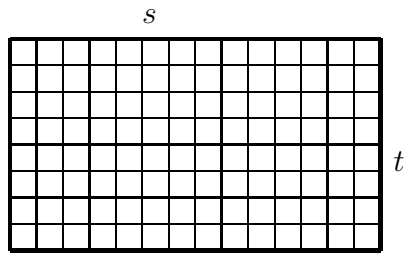
$$
{}
$$
\bigskip
\begin{center}
\be
\label{block}
\begin{picture}(70,35)
\put(50,96){$s$}
{\linethickness{.5mm}
\put(00,90){\line(1,0){140}}%
\put(00,10){\line(1,0){140}}%
\put(140,10){\line(0,1){80}}%
\put(00,10){\line(0,1){80}}%
}%
\put(00,70){\line(1,0){140}}
\put(00,80){\line(1,0){140}}
\put(00,60){\line(1,0){140}}
\put(00,50){\line(1,0){140}}
\put(00,40){\line(1,0){140}}
\put(00,30){\line(1,0){140}}
\put(00,20){\line(1,0){140}}
\put(10,10.0){\line(0,1){80}}
\put(20,10.0){\line(0,1){80}}
\put(30,10.0){\line(0,1){80}}
\put(40,10.0){\line(0,1){80}}
\put(50,10.0){\line(0,1){80}}
\put(60,10.0){\line(0,1){80}}
\put(70,10.0){\line(0,1){80}}
\put(80,10.0){\line(0,1){80}}
\put(90,10.0){\line(0,1){80}}
\put(100,10.0){\line(0,1){80}}
\put(110,10.0){\line(0,1){80}}
\put(120,10.0){\line(0,1){80}}
\put(130,10.0){\line(0,1){80}}
\put(145,40){$t$}%
\end{picture}
\ee
\end{center}
In other words, a block is a Young diagram composed of
$t$ rows of equal length $s$
(equivalently, from $s$ columns of equal height $t$).
A general Young diagram is a combination of blocks
with decreasing lengths (equivalently, heights)
\newpage
$${}$$
\be
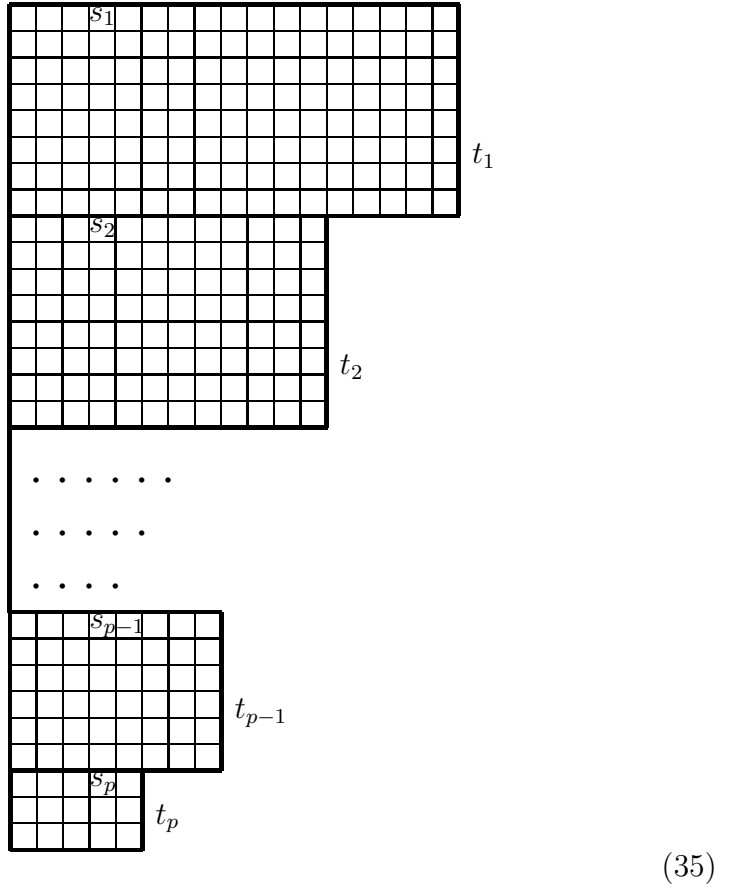
\label{blockd}
\bigskip
\begin{picture}(85,300)%
\begin{picture}(109,300)%
{\linethickness{.500mm}
\put(30,34){$s_{p}$}
\put(50,10){\line(0,1){30}}
\put(00,10){\line(1,0){50}}
\put(00,10){\line(0,1){30}}
\put(00,40){\line(1,0){50}}}
\put(.0,40){\line(1,0){50}}
\put(.0,30){\line(1,0){50}}
\put(.0,20){\line(1,0){50}}
\put(10,10){\line(0,1){30}}
\put(20,10){\line(0,1){30}}
\put(30,10){\line(0,1){30}}
\put(40,10){\line(0,1){30}}
\put(55,20){$t_{p}$}
\end{picture}
\begin{picture}(125,90)(113,-30)%
{\linethickness{.500mm}
\put(30,64){$s_{p-1}$}
\put(80,10){\line(0,1){60}}
\put(00,10){\line(1,0){80}}
\put(00,10){\line(0,1){60}}
\put(00,70){\line(1,0){80}}}
\put(.0,60){\line(1,0){80}}
\put(.0,50){\line(1,0){80}}
\put(.0,40){\line(1,0){80}}
\put(.0,30){\line(1,0){80}}
\put(.0,20){\line(1,0){80}}
\put(10,10){\line(0,1){60}}
\put(20,10){\line(0,1){60}}
\put(30,10){\line(0,1){60}}
\put(40,10){\line(0,1){60}}
\put(50,10){\line(0,1){60}}
\put(60,10){\line(0,1){60}}
\put(70,10){\line(0,1){60}}
\put(85,30){$t_{p-1}$}
\end{picture}
\begin{picture}(75,75)(242,-90)%
{\linethickness{.5mm}
\put(00,10){\line(0,1){70}}}
\put(10,20){\circle*{2}}
\put(20,20){\circle*{2}}
\put(30,20){\circle*{2}}
\put(40,20){\circle*{2}}
\put(50,40){\circle*{2}}
\put(40,40){\circle*{2}}
\put(30,40){\circle*{2}}
\put(10,40){\circle*{2}}
\put(20,40){\circle*{2}}
\put(60,60){\circle*{2}}
\put(50,60){\circle*{2}}
\put(40,60){\circle*{2}}
\put(30,60){\circle*{2}}
\put(10,60){\circle*{2}}
\put(20,60){\circle*{2}}
\end{picture}
\begin{picture}(125,95)(321,-160)%
\put(30,84){$s_2$}%
{\linethickness{.500mm}
\put(120,10){\line(0,1){80}}
\put(00,90){\line(1,0){120}}
\put(00,10){\line(1,0){120}}
\put(00,10){\line(0,1){80}}}
\put(00,70){\line(1,0){120}}
\put(00,80){\line(1,0){120}}
\put(00,60){\line(1,0){120}}
\put(00,50){\line(1,0){120}}
\put(00,40){\line(1,0){120}}
\put(00,30){\line(1,0){120}}
\put(00,20){\line(1,0){120}}
\put(10,10.0){\line(0,1){80}}
\put(20,10.0){\line(0,1){80}}
\put(30,10.0){\line(0,1){80}}
\put(40,10.0){\line(0,1){80}}
\put(50,10.0){\line(0,1){80}}
\put(60,10.0){\line(0,1){80}}
\put(70,10.0){\line(0,1){80}}
\put(80,10.0){\line(0,1){80}}
\put(90,10.0){\line(0,1){80}}
\put(100,10.0){\line(0,1){80}}
\put(110,10.0){\line(0,1){80}}
\put(125,30){$t_2$}
\end{picture}
\begin{picture}(175,85)(450,-240)%
\put(30,84){$s_1$}
{\linethickness{.500mm}
\put(00,90){\line(1,0){170}}%
\put(00,10){\line(1,0){170}}%
\put(00,10){\line(0,1){80}}%
\put(170,10){\line(0,1){80}}}%
\put(00,70){\line(1,0){170}}
\put(00,80){\line(1,0){170}}
\put(00,60){\line(1,0){170}}
\put(00,50){\line(1,0){170}}
\put(00,40){\line(1,0){170}}
\put(00,30){\line(1,0){170}}
\put(00,20){\line(1,0){170}}
\put(10,10.0){\line(0,1){80}}
\put(20,10.0){\line(0,1){80}}
\put(30,10.0){\line(0,1){80}}
\put(40,10.0){\line(0,1){80}}
\put(50,10.0){\line(0,1){80}}
\put(60,10.0){\line(0,1){80}}
\put(70,10.0){\line(0,1){80}}
\put(80,10.0){\line(0,1){80}}
\put(90,10.0){\line(0,1){80}}
\put(100,10.0){\line(0,1){80}}
\put(110,10.0){\line(0,1){80}}
\put(120,10.0){\line(0,1){80}}
\put(130,10.0){\line(0,1){80}}
\put(140,10.0){\line(0,1){80}}
\put(150,10.0){\line(0,1){80}}
\put(160,10.0){\line(0,1){80}}%
\put(175,30){$t_1$}%
\end{picture}
\end{picture}
\ee

In these terms, a Young diagram $Y[(s_i,t_i) ]$
is described by a set of pairs of positive integers
$(s_i , t_i)$ with $s_1 > s_2 > s_3 > \ldots s_p >0$
and arbitrary $t_i$ such that
\be
\sum_{i=1}^{p} t_i \leq \half (d-1)\,.
\ee
In other words, $s_1$ is the maximal row length in the
Young diagram, while $t_1$ is the number of rows of
the length $s_1$. $s_2$ is the maximal row length
of the remaining rows and $t_2$ is the number of rows of length
$s_2$. Note that an elementary block $Y[(s,t)]$
is described in these terms
by a single pair of integers $(s ,t) $.

Let us now address the question what is the result of a dimensional
reduction to one dimension less of a general
diagram $Y[(s_i ,t_i )]$.
Every cell can be identified with some
vector index. It can either be aligned along $n^a$ or along the
$d-2$ perpendicular directions. In the first case we
cancel a cell, while in the second case we keep it. There cannot
be more than $s_1$ indices along $n^a$ because symmetrization
with respect to more than $s$ indices gives identically zero
by the definition of a Young diagram. But any number of indices
from 0 to $s_1$ can be chosen to take the extra value $d-1$. Therefore
any number of cells from 0 to $s_1$ can be canceled. Of course
only such cancelings are allowed that   result in a Young diagram.
(If not,  any resulting tensor is
identically zero.)

Let us consider some examples.

First consider a representation
(e.g. tensor $T$) described by the Young diagram
$Y[(s,t)]$ which is itself an elementary block. Since components
of tensors along $n_a$ are automatically symmetrized
because the tensor $N^{a_1 a_2 \ldots a_k} =
n^{a_1}n^{a_2}\ldots n^{a_k}$ is totally symmetric,
we can use the symmetry properties of the Young diagrams to reduce
any contraction with $N^{a_1 a_2 \ldots a_k}$
of the original tensor to a contraction of
$N^{a_1 a_2 \ldots a_k}$ with the bottom row of the block.
As a result, dimensional reduction will lead to a number of
tensors resulting from cutting an arbitrary
number of cells in the bottom row of the block; every tensor
appears once (note that the condition that the tensor is
traceless does not affect this analysis since the reduced
$o(d-2)$ tensors are also assumed to be traceless).

\be
\begin{picture}(240,85)
\put(30,83){$s_1$}
{\linethickness{.5mm}
\put(00,90){\line(1,0){170}}%
\put(00,10){\line(1,0){140}}%
\put(140,10){\line(0,1){10}}%
\put(0,20){\line(1,0){170}}%
\put(00,10){\line(0,1){80}}%
\put(170,20){\line(0,1){70}}}%
\put(00,70){\line(1,0){170}}
\put(00,80){\line(1,0){170}}
\put(00,60){\line(1,0){170}}
\put(00,50){\line(1,0){170}}
\put(00,40){\line(1,0){170}}
\put(00,30){\line(1,0){170}}
\put(00,20){\line(1,0){170}}
\put(10,10.0){\line(0,1){80}}
\put(20,10.0){\line(0,1){80}}
\put(30,10.0){\line(0,1){80}}
\put(40,10.0){\line(0,1){80}}
\put(50,10.0){\line(0,1){80}}
\put(60,10.0){\line(0,1){80}}
\put(70,10.0){\line(0,1){80}}
\put(80,10.0){\line(0,1){80}}
\put(90,10.0){\line(0,1){80}}
\put(100,10.0){\line(0,1){80}}
\put(110,10.0){\line(0,1){80}}
\put(120,10.0){\line(0,1){80}}
\put(130,10.0){\line(0,1){80}}
\put(140,20.0){\line(0,1){70}}
\put(150,20.0){\line(0,1){70}}
\put(160,20.0){\line(0,1){70}}%
\put(155,12){$n_1$}
\put(175,40){$t_1-1$}%
\end{picture}
\ee
In other words, the dimensional reduction of the
block  $Y[(s,t)]$ gives rise to the following representations of
$o(d-2)$: $Y[(s,t)]$ (all indices are orthogonal to $n^a$),
$Y[(s,t-1)]$ (a maximal possible number $s$ of indices is
contracted) and all diagrams  $Y[(s,t-1);(s_1 ,1)]$ which consist
of two blocks with the bottom block having an arbitrary length
$0<s_1< s$ and height 1.

Now, consider a representation $T$ described by a Young
diagram $Y[(s_1 ,t_1 );(s_2 ,t_2 )]$ composed from two blocks.
\be
\begin{picture}(110,185)             
\put(30,84){$s_2$}%
{\linethickness{.500mm}
\put(120,10){\line(0,1){80}}
\put(00,90){\line(1,0){120}}
\put(00,10){\line(1,0){120}}
\put(00,10){\line(0,1){80}}}
\put(00,70){\line(1,0){120}}
\put(00,80){\line(1,0){120}}
\put(00,60){\line(1,0){120}}
\put(00,50){\line(1,0){120}}
\put(00,40){\line(1,0){120}}
\put(00,30){\line(1,0){120}}
\put(00,20){\line(1,0){120}}
\put(10,10.0){\line(0,1){80}}
\put(20,10.0){\line(0,1){80}}
\put(30,10.0){\line(0,1){80}}
\put(40,10.0){\line(0,1){80}}
\put(50,10.0){\line(0,1){80}}
\put(60,10.0){\line(0,1){80}}
\put(70,10.0){\line(0,1){80}}
\put(80,10.0){\line(0,1){80}}
\put(90,10.0){\line(0,1){80}}
\put(100,10.0){\line(0,1){80}}
\put(110,10.0){\line(0,1){80}}
\put(125,30){$t_2$}
\end{picture}
\begin{picture}(160,85)(110,-80)%
\put(30,84){$s_1$}
{\linethickness{.500mm}
\put(00,90){\line(1,0){170}}%

\put(00,10){\line(1,0){170}}%
\put(00,10){\line(0,1){80}}%
\put(170,10){\line(0,1){80}}}%
\put(00,70){\line(1,0){170}}
\put(00,80){\line(1,0){170}}
\put(00,60){\line(1,0){170}}
\put(00,50){\line(1,0){170}}
\put(00,40){\line(1,0){170}}
\put(00,30){\line(1,0){170}}
\put(00,20){\line(1,0){170}}
\put(10,10.0){\line(0,1){80}}
\put(20,10.0){\line(0,1){80}}
\put(30,10.0){\line(0,1){80}}
\put(40,10.0){\line(0,1){80}}
\put(50,10.0){\line(0,1){80}}
\put(60,10.0){\line(0,1){80}}
\put(70,10.0){\line(0,1){80}}
\put(80,10.0){\line(0,1){80}}
\put(90,10.0){\line(0,1){80}}
\put(100,10.0){\line(0,1){80}}
\put(110,10.0){\line(0,1){80}}
\put(120,10.0){\line(0,1){80}}
\put(130,10.0){\line(0,1){80}}
\put(140,10.0){\line(0,1){80}}
\put(150,10.0){\line(0,1){80}}
\put(160,10.0){\line(0,1){80}}%
\put(175,30){$t_1$}%
\end{picture}
\ee
\bigskip

Again, dimensional reduction means that one can cut some
cells from the bottom rows of the upper and lower blocks
(all cuts inside a block are equivalent by the properties
of the Young diagrams to cutting its bottom line). But now,
one cannot cut an arbitrary number of boxes in the top
block because the cut line cannot be shorter than the
length of the second block $s_2$. (Such  tensors
vanish identically.) One can, however, take away an arbitrary
number of cells from the bottom line of the second box. The
rule therefore is: take away an arbitrary number $n_1$ such that
$ s_1 -s_2\geq n_1 \geq 0$ from the bottom line of the top block
and take away an arbitrary number $n_2$ such that
$ s_2 \geq n_2 \geq 0 $ of cells from the bottom line of the bottom
block
\bigskip
\
\be 
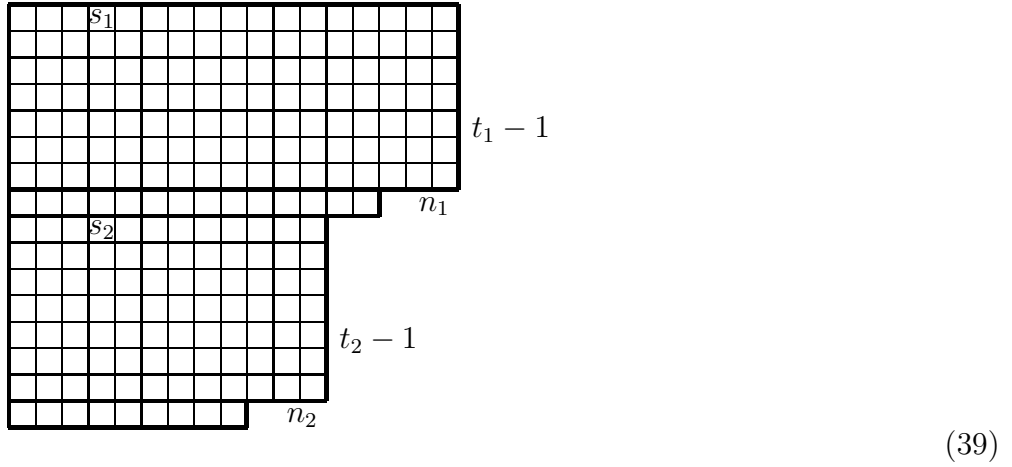
\begin{picture}(125,185)             
\put(30,83){$s_2$}%
{\linethickness{.5mm}
\put(120,20){\line(0,1){70}}
\put(00,90){\line(1,0){120}}
\put(00,10){\line(1,0){90}}
\put(90,10){\line(0,1){10}}
\put(00,20){\line(1,0){120}}
\put(00,10){\line(0,1){80}}}
\put(00,70){\line(1,0){120}}
\put(00,80){\line(1,0){120}}
\put(00,60){\line(1,0){120}}
\put(00,50){\line(1,0){120}}
\put(00,40){\line(1,0){120}}
\put(00,30){\line(1,0){120}}
\put(00,20){\line(1,0){120}}
\put(10,10.0){\line(0,1){80}}
\put(20,10.0){\line(0,1){80}}
\put(30,10.0){\line(0,1){80}}
\put(40,10.0){\line(0,1){80}}
\put(50,10.0){\line(0,1){80}}
\put(60,10.0){\line(0,1){80}}
\put(70,10.0){\line(0,1){80}}
\put(80,10.0){\line(0,1){80}}
\put(90,20){\line(0,1){70}}
\put(100,20.0){\line(0,1){70}}
\put(110,20.0){\line(0,1){70}}
\put(105,12){$n_2$}
\put(125,40){$t_2 -1$}
\end{picture}
\begin{picture}(175,85)(125,-80)%
\put(30,83){$s_1$}
{\linethickness{.5mm}
\put(00,90){\line(1,0){170}}%
\put(00,10){\line(1,0){140}}%
\put(140,10){\line(0,1){10}}%
\put(00,20){\line(1,0){170}}%
\put(00,10){\line(0,1){80}}%
\put(170,20){\line(0,1){70}}}%
\put(00,70){\line(1,0){170}}
\put(00,80){\line(1,0){170}}
\put(00,60){\line(1,0){170}}
\put(00,50){\line(1,0){170}}
\put(00,40){\line(1,0){170}}
\put(00,30){\line(1,0){170}}
\put(00,20){\line(1,0){170}}
\put(10,10.0){\line(0,1){80}}
\put(20,10.0){\line(0,1){80}}
\put(30,10.0){\line(0,1){80}}
\put(40,10.0){\line(0,1){80}}
\put(50,10.0){\line(0,1){80}}
\put(60,10.0){\line(0,1){80}}
\put(70,10.0){\line(0,1){80}}
\put(80,10.0){\line(0,1){80}}
\put(90,10.0){\line(0,1){80}}
\put(100,10.0){\line(0,1){80}}
\put(110,10.0){\line(0,1){80}}
\put(120,10.0){\line(0,1){80}}
\put(130,10.0){\line(0,1){80}}
\put(140,20.0){\line(0,1){70}}
\put(150,20.0){\line(0,1){70}}
\put(160,20.0){\line(0,1){70}}%
\put(155,12){$n_1$}
\put(175,40){$t_1 -1$}%
\end{picture}
\ee
Note that with this
prescription we have $n_1 + n_2 \leq s_1$ in accordance with the
general argument that one cannot cut a number of cells exceeding
the maximal row length in the Young diagram. The pattern of the
dimensionally reduced representation therefore consists of
four-block diagrams
$Y[(s_1 ,t_1 -1); (s_1^\prime ,1); (s_2 , t_2 -1); (s_2^\prime ,1)]$
with arbitrary integers $s_1^\prime$ and $s_2^\prime$ such that
\be
s_1 > s_1^\prime > s_2 > s_2^\prime >0
\ee
and their degenerate versions described by the
 three-block diagrams \\ $Y[(s_1 , t_1 -1);(s_2 , t_2 ); (s_2^\prime , 1)]$,
$Y[(s_1 , t_1 );(s_2 , t_2 -1); (s_2^\prime , 1)]$,
$Y[(s_1 , t_1 -1 );(s_1^\prime , 1);(s_2 , t_2 )]$,
$Y[(s_1 , t_1 -1 );(s_1^\prime , 1);(s_2 , t_2 -1)]$ and two-block diagrams
$Y[(s_1 , t_1 );(s_2 , t_2 )]$, \\ $Y[(s_1 , t_1 -1);(s_2 , t_2 )]$,
$Y[(s_1 , t_1 );(s_2 , t_2 -1)]$,  $Y[(s_1 , t_1 -1);(s_2 , t_2 +1)]$.

Analogously one proceeds for Young diagrams built from a larger
number of blocks. The final result is that the dimensional reduction
of a general Young diagram to one dimension less consists of the Young
diagrams of the form (every diagram appears once; see e.g.
\cite{barut} and references therein):
\newpage
$${}$$
\be
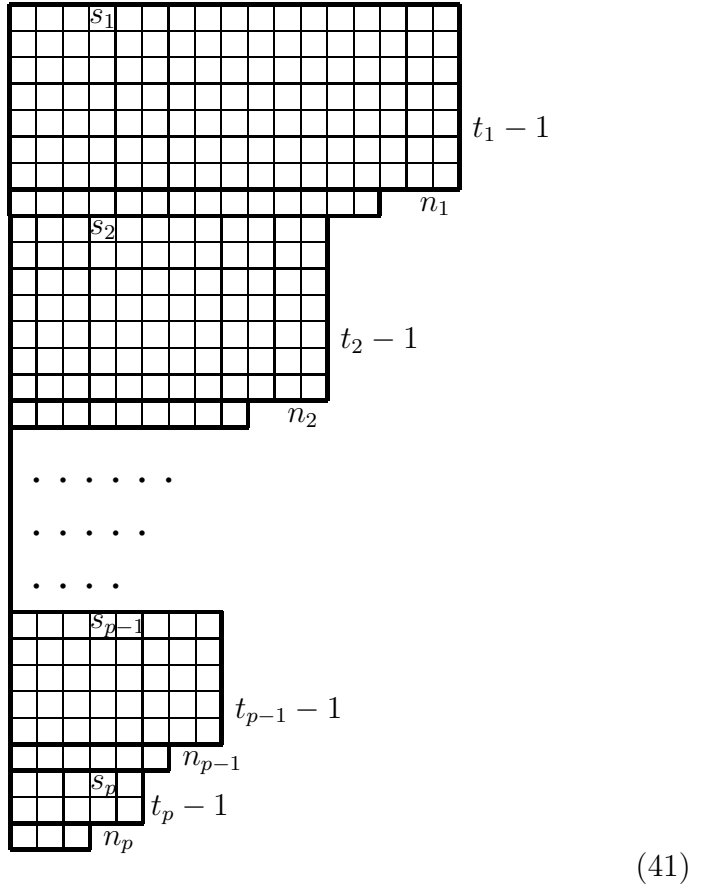
\begin{picture}(65,300)%
\begin{picture}(112,300)%
{\linethickness{.5mm}
\put(30,33){$s_{p}$}
\put(50,20){\line(0,1){20}} \put(00,10){\line(1,0){30}}
\put(00,20){\line(1,0){50}} \put(30,10){\line(0,1){10}}
\put(00,10){\line(0,1){30}} \put(00,40){\line(1,0){50}}}
\put(.0,40){\line(1,0){50}} \put(.0,30){\line(1,0){50}}
\put(.0,20){\line(1,0){50}} \put(10,10){\line(0,1){30}}
\put(20,10){\line(0,1){30}} \put(30,20){\line(0,1){20}}
\put(40,20){\line(0,1){20}} \put(35,12){$n_{p}$}
\put(53,23){$t_{p}-1$}
\end{picture}
\begin{picture}(125,90)(116,-30)%
{\linethickness{.500mm}
\put(30,64){$s_{p-1}$} \put(80,20){\line(0,1){50}}
\put(00,10){\line(1,0){60}} \put(60,10){\line(0,1){10}}
\put(0,20){\line(1,0){80}} \put(00,10){\line(0,1){60}}
\put(00,70){\line(1,0){80}}} \put(.0,60){\line(1,0){80}}
\put(.0,50){\line(1,0){80}} \put(.0,40){\line(1,0){80}}
\put(.0,30){\line(1,0){80}} \put(.0,20){\line(1,0){80}}
\put(10,10){\line(0,1){60}} \put(20,10){\line(0,1){60}}
\put(30,10){\line(0,1){60}} \put(40,10){\line(0,1){60}}
\put(50,10){\line(0,1){60}} \put(60,20){\line(0,1){50}}
\put(70,20){\line(0,1){50}} \put(65,12){$n_{p-1}$}
\put(85,30){$t_{p-1}-1$} %
\end{picture}
\begin{picture}(75,75)(245,-90)%
{\linethickness{.5mm} \put(00,10){\line(0,1){70}}}
\put(10,20){\circle*{2}} \put(20,20){\circle*{2}}
\put(30,20){\circle*{2}} \put(40,20){\circle*{2}}
\put(50,40){\circle*{2}} \put(40,40){\circle*{2}}
\put(30,40){\circle*{2}} \put(10,40){\circle*{2}}
\put(20,40){\circle*{2}} \put(60,60){\circle*{2}}
\put(50,60){\circle*{2}} \put(40,60){\circle*{2}}
\put(30,60){\circle*{2}} \put(10,60){\circle*{2}}
\put(20,60){\circle*{2}} \end{picture}
\begin{picture}(125,95)(324,-160)%
\put(30,83){$s_2$}%
{\linethickness{.5mm}
\put(120,20){\line(0,1){70}}
\put(00,90){\line(1,0){120}} \put(00,10){\line(1,0){90}}
\put(90,10){\line(0,1){10}} \put(0,20){\line(1,0){120}}
\put(00,10){\line(0,1){80}}} \put(00,70){\line(1,0){120}}
\put(00,80){\line(1,0){120}} \put(00,60){\line(1,0){120}}
\put(00,50){\line(1,0){120}} \put(00,40){\line(1,0){120}}
\put(00,30){\line(1,0){120}} \put(00,20){\line(1,0){120}}
\put(10,10.0){\line(0,1){80}} \put(20,10.0){\line(0,1){80}}
\put(30,10.0){\line(0,1){80}} \put(40,10.0){\line(0,1){80}}
\put(50,10.0){\line(0,1){80}} \put(60,10.0){\line(0,1){80}}
\put(70,10.0){\line(0,1){80}} \put(80,10.0){\line(0,1){80}}
\put(90,20){\line(0,1){70}} \put(100,20.0){\line(0,1){70}}
\put(110,20.0){\line(0,1){70}} \put(105,12){$n_2$}
\put(125,40){$t_2-1$}
\end{picture}
\begin{picture}(175,85)(453,-240)%
\put(30,83){$s_1$}
{\linethickness{.5mm}
\put(00,90){\line(1,0){170}}%
\put(00,10){\line(1,0){140}}%
\put(140,10){\line(0,1){10}}%
\put(0,20){\line(1,0){170}}%
\put(00,10){\line(0,1){80}}%
\put(170,20){\line(0,1){70}}}%
\put(00,70){\line(1,0){170}}
\put(00,80){\line(1,0){170}} \put(00,60){\line(1,0){170}}
\put(00,50){\line(1,0){170}} \put(00,40){\line(1,0){170}}
\put(00,30){\line(1,0){170}} \put(00,20){\line(1,0){170}}
\put(10,10.0){\line(0,1){80}}
\put(20,10.0){\line(0,1){80}} \put(30,10.0){\line(0,1){80}}
\put(40,10.0){\line(0,1){80}} \put(50,10.0){\line(0,1){80}}
\put(60,10.0){\line(0,1){80}} \put(70,10.0){\line(0,1){80}}
\put(80,10.0){\line(0,1){80}} \put(90,10.0){\line(0,1){80}}
\put(100,10.0){\line(0,1){80}} \put(110,10.0){\line(0,1){80}}
\put(120,10.0){\line(0,1){80}} \put(130,10.0){\line(0,1){80}}
\put(140,20.0){\line(0,1){70}} \put(150,20.0){\line(0,1){70}}
\put(160,20.0){\line(0,1){70}}%
\put(155,12){$n_1$}
\put(175,40){$t_1-1$}%
\end{picture}
\end{picture}
\ee
One is allowed to take away any numbers $n_i$ of cells from the
$i^{th}$ block provided that
\be
\label{ni}
 0\leq  n_i \leq s_i - s_{i+1}
\ee
(with the convention that $s_j$ corresponding to the ``next to last''
block equals zero).

Let us now  formulate the final result concerning
a pattern of flat space massless fields that admit a unitary
deformation to $AdS_d$.

{\bf Conjecture.}
{\it Consider a $AdS_d$ massless field characterized by a
diagram (\ref{blockd}). Take away any number of
cells $n_i$  satisfying the conditions (\ref{ni}) of the bottom lines of all
blocks except for the top one, i.e. require $n_1 =0$.  Any diagram that appears
as a result describes some irreducible representation of the massless little
algebra  $o(d-2)$ corresponding to some flat space massless field that should be
present in the full set compatible with the $AdS_d$ geometry and unitarity.}

Note that massless fields corresponding to arbitrary
irreducible representations of flat space massless
little algebra $o(d-2)$ were considered in \cite{labas}.

A few comments are now in order.

The role of the upper block is singled out by the unitarity
condition: only singular vectors corresponding to contractions
of $t_+^a$ to the upper block have maximal energies and
describe unitary representations \cite{met95}. Therefore canceling
out a box from the upper block corresponds to pure gauge
(i.e. singular vector) components that decouple.
Non-unitary (i.e. ghost containing) sets of fields can be obtained
by a similar procedure with one of the lower blocks remaining untouched
instead of the top one as in the unitary case.

For Young diagrams being themselves elementary blocks
(i.e. $t_i=0$ for $i>0$) only one $o(d-2)$ representation
appears, described by the same block. This means that elementary
block massless fields in $AdS_d$ classify according to
irreducible representations
of $o(d-2)$ as in the flat case (i.e. no additional massless fields
should be added to deform to $AdS_d$). In fact all examples of
massless fields that appear in supergravity and
low energy string theory
are described by elementary blocks (specifically,
either by single rows, or by single columns). That is why the
phenomenon discussed in this paper was not observed before.
Note also that for the well studied case of lower
dimensions $d\leq 4$, only block-type massless representations are
nontrivial (propagating) and therefore this phenomenon does not occur
either.

The spectrum of flat-space massless fields to which an elementary
$AdS$ massless field decomposes is non degenerate, i.e. all the
representations of $o(d-2)$ are pairwise different.

Some standard massless (gauge) fields
may be needed as ingredients of $AdS$ massless fields with
nontrivial diagrams. For example, for the representation $Y[(2,1),(1,1)]$
a graviton-type flat space massless field $Y[(2,1)]$ will be present
(see example in section \ref{Gauge Symmetries in $AdS_d$}). It is
tempting to speculate that this may correspond to a nontrivial
deformation of gravity to the $AdS$ geometry in the presence of other
fields.
Analogously one can find a totally symmetric spin  $s_1\geq 2$
field corresponding to the diagram $Y[(s_1,1)]$ among the fields
resulting from the
decomposition of the $AdS_d$ field
$Y[(s_1,1);(s_2,1)]$, $0 \leq s_2\leq s_1$. This is to say that
$AdS_d$ massless field
corresponding to
$Y[(s_1,1);(s_2,1)]$ decomposes into the following
irreps of $o(d-2)$ algebra
\be
Y[(s_1,1);(s_2,1)]_{AdS}\rightarrow Y[(s_1,1)]\oplus
\sum_{s=1}^{s_2} \oplus Y[(s_1,1);(s,1)],
\ee
where each term under summation appears just once.

An important consequence of the analysis of this section is that
totally antisymmetric gauge tensors
(i.e. differential forms) corresponding to the diagrams Y[(1,t)]
can never appear as a result of a decomposition of a certain irreducible
(unitary)
$AdS_d$ massless field in the flat limit. The space of
differential forms (including the spin one gauge fields)
therefore is closed with respect to the deformation
to $AdS_d$.

\section{Field Theoretical Example}

Now let us explain what happens from the field-theoretic
perspective. In fact, it has been observed already in
\cite{met95} at the level of
equations of motion in the Lorentz gauge that some of
the redundant gauge symmetries expected in the flat-space description
are absent in the AdS case. Here we analyze the problem at
the Lagrangian level focusing on the explicit comparison
with the flat-space limit.

\subsection{Flat Space}
\label{Flat Space}

Let us consider the
simplest nontrivial example of a massless field having the
symmetry properties of a non-block diagram with three cells
$Y[(2,1);(1,1)]$

\medskip
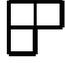
\be
\label{YD21}
\begin{picture}(130,45)
{\linethickness{.500mm}
\put(00,20){\line(1,0){20}}%
\put(00,30){\line(1,0){20}}%
\put(00,10){\line(1,0){10}}%
\put(00,10){\line(0,1){20}}%
\put(10,10){\line(0,1){10}}
\put(20,20){\line(0,1){10}}}%
\put(10,20){\line(0,1){10}}%
\end{picture}
\ee

In flat space the massless field of this symmetry type
was described in \cite{AKO}. To begin with, let us reformulate
the results of these authors in a somewhat different, although
equivalent way. We take the representation with the field
$\Phi_{m_1 m_2,n}$ being  a symmetric tensor in $m_1$ and $m_2$,
satisfying the condition that full symmetrization with respect
to all three indices gives zero
\be
\label{fulsym}
\Phi_{\{m_1 m_2, m_3\}} =0\,.
\ee
(The authors of \cite{AKO} used an equivalent representation
with explicit antisymmetry in two indices).
The Lagrangian can be chosen to be of the form
\begin{eqnarray}
L &=& \frac{1}{2}\Phi_{m_1m_2,n}\Box \Phi^{m_1m_2,n}
-\Phi_{mm_1,n}\partial^{m_1}\partial_{m_2}\Phi^{mm_2,n}
-\frac{1}{2}\Phi_{m_1m_2,n_1}\partial^{n_1}
\partial_{n_2}\Phi^{m_1m_2,n_2}
\nonumber\\
&-&\frac{3}{4}\Phi^{m_1}{}_{m_1,n}\Box\Phi_{m_2}{}^{m_2,n}
+\frac{3}{4}\Phi^{m}{}_{m,n}\partial_{m_1}\partial_{m_2}\Phi^{m_1m_2,n}
+\frac{3}{4}\Phi^{m_1m_2,n}\partial_{m_1}\partial_{m_2}\Phi^{m}{}_{m,n}
\nonumber\\
\label{minlagten}
&+&
\frac{3}{4}\Phi^{m_1}{}_{m_1,n_1}\partial^{n_1}\partial_{n_2}
\Phi_{m_2}{}^{m_2,n_2}\,.
\end{eqnarray}
The corresponding action  is invariant under the gauge transformations
\begin{eqnarray}
\label{flagtr1}
&&
\delta_{as} \Phi^{m_1m_2,n}
=\frac{1}{2}(\partial^{m_1}\Lambda_{as}^{m_2 n}
+\partial^{m_2}\Lambda_{as}^{m_1 n})\,,
\\
\label{flagtr2}
&&
\delta_{sym}\Phi^{m_1m_2,n}
=\frac{1}{2}(\partial^{m_1}\Lambda_{sym}^{m_2n}
+\partial^{m_2}\Lambda_{sym}^{m_1n})
-\partial^n\Lambda_{sym}^{m_1m_2}\,
\end{eqnarray}
with antisymmetric gauge parameter $\Lambda_{as}^{m n} (x)$
and symmetric gauge parameter $\Lambda_{sym}^{m n} (x)$,
\be
\Lambda_{as}^{m n} (x) =- \Lambda_{as}^{ n m} (x) \,,\qquad
\Lambda_{sym}^{m n} (x) =\Lambda_{sym}^{ n m} (x) \,.
\ee

In \cite{AKO} it was proved that the Lagrangian
(\ref{minlagten}) describes a physical massless field in flat space
corresponding to the irreducible representation $Y[(2,1);(1,1)]$
of the massless little algebra $o(d-2)$. Because this Lagrangian
is fixed by the gauge transformations one can say that it is the gauge
invariance with respect to the both gauge transformations (\ref{flagtr1})
and (\ref{flagtr2}) that ensures irreducibility of the massless
field.

Let us now introduce a notation that simplifies computation.
In curved space-time it is convenient to
use fiberwise fields
\begin{equation}\label{tantar}
\Phi^{m_1m_2m_3}
\equiv e_{\underline{m}_1}{}^{m_1}e_{\underline{m}_2}{}^{m_2}
e_{\underline{m}_3}{}^{m_3}
\Phi^{\underline{m}_1\underline{m}_2\underline{m}_3}\,,
\end{equation}
where $e_{\underline{m}}{}^m$ is the vielbein of
an appropriate space-time (e.g., Minkowski or
AdS) \footnote{Tangent space (fiber) indices $m,n$ and target space
(base) indices ${\underline{m}},{\underline{n}}$ take the values
$0,1,\ldots\,d-1$.}.

It is most convenient to formulate the action in terms of the following
Fock-type generating function \cite{LV}
\begin{equation}
\label{fockv}
|\Phi\rangle
=\frac{1}{\sqrt{2}}
\Phi_{m_1m_2,n}\alpha_1^{m_1}\alpha_1^{m_2}\alpha_2^n|0\rangle\,,
\end{equation}
where $\alpha_A^m$ and $\bar{\alpha}_B^n$ are auxiliary creation and
annihilation operators
\begin{equation}\label{antosc}
[\bar{\alpha}_A^m,\alpha_B^n]=\eta^{mn}\delta_{AB}\,,
\qquad
[\alpha_A^m,\alpha_B^n]=0\,,
\qquad
[\bar{\alpha}_A^m,\bar{\alpha}_B^n]=0,
\ee
and $|0\rangle$ is a Fock vacuum
\be
\bar{\alpha}_A^m|0\rangle=0 \,.
\end{equation}
The indices $A,B,C,E = 1,2$ label two sets of oscillators.

The fact that we deal with the Young diagram (\ref{YD21})
is equivalent to imposing the following constraints on the
 generating function $|\Phi\rangle$
\begin{eqnarray}
\label{n11}
&&
N_{11}|\Phi\rangle=2|\Phi\rangle\,,
\\
\label{n22}
&&
N_{22}|\Phi\rangle=|\Phi\rangle\,,
\\
\label{n12}
&&
N_{12}|\Phi\rangle=0\,,
\end{eqnarray}
where we use the notation
\begin{equation}\label{defnij}
N_{AB}\equiv \alpha_A^m\bar{\alpha}_{Bm}\,,
\qquad
P_{AB}\equiv \alpha_A^m\alpha_{Bm}\,,
\qquad
\bar{P}_{AB}\equiv \bar{\alpha}_A^m\bar{\alpha}_{Bm}\,.
\end{equation}
The constraints (\ref{n11}) and (\ref{n22}) tell us
that the oscillators $\alpha_1^m$ and $\alpha_2^m$ occur twice
and once, respectively, on the right hand side of eq.(\ref{fockv}).
The constraint (\ref{n12}) is equivalent to the condition
(\ref{fulsym}).

The Lorentz covariant derivative for the representation
$|\Phi\rangle$ takes the form
\begin{equation}\label{lorspiope}
D_{\underline{m}}\equiv
\partial_{\underline{m}}
+\frac{1}{2}\omega_{\underline{m}}{}_{mn}M^{mn}\,,
\qquad
M^{mn}
=\sum_{A=1,2}
(\alpha_A^m\bar{\alpha}_A^n-\alpha_A^n\bar{\alpha}_A^m)\,,
\end{equation}
where $\omega_{\underline{m}}{}^{mn}$ is the Lorentz connection of
space, while $M^{mn}$ forms a representation of the Lorentz algebra
$so(d-1,1)$.  In the sequel we will often use  the notation

\begin{equation}\label{conv}
D_A\equiv \alpha_A{}^m D_m\,,
\qquad
\bar{D}_A\equiv \bar{\alpha}_A{}^m D_m\,,
\qquad
D_m\equiv e_m{}^{\underline{m}} D_{\underline{m}}\,.
\end{equation}

The flat-space Lagrangian (\ref{minlagten}) now takes the form

\begin{equation}\label{minlag}
L
=\frac{1}{2}\langle\Phi|
\Box-D_1 \bar{D}_1 -D_2 \bar{D}_2
-\frac{3}{4} P_{11}\Box\bar{P}_{11}
+\frac{3}{4} (P_{11}\bar{D}_1^2 + D_1^2\bar{P}_{11}
+P_{11}D_2\bar{D}_2 \bar{P}_{11})|\Phi \rangle\,,
\end{equation}
where
$D_A$, $\bar{D}_B$ and $\Box = D^m D_m $ are defined via
(\ref{lorspiope}) and (\ref{conv}) with the flat space
vielbein and Lorentz connection
$$
e_{\underline{m}}{}^n=\delta_{\underline{m}}^n\,,
\qquad
\omega_{\underline{m}}{}^{kn}=0\,.
$$
The Lagrangian (\ref{minlag}) is invariant under two gauge
symmetries  generated by the antisymmetric gauge parameter
$\Lambda_{as}^{mn}$ and the symmetric gauge parameter
$\Lambda_{sym}^{mn}$ which can be conveniently described as
Fock vectors

$$
|\Lambda_{as}\rangle
=\frac{1}{\sqrt{2}}\Lambda_{as\,mn}\alpha_1^m\alpha_2^n|0\rangle\,,
\qquad
|\Lambda_{sym}\rangle
=\frac{1}{\sqrt{2}}\Lambda_{sym\,mn}\alpha_1^m\alpha_1^n|0\rangle\,.
$$
Now the gauge transformations (\ref{flagtr1}), (\ref{flagtr2})
take the form

\begin{eqnarray}
\label{mingtras}
&&
\delta_{as}|\Phi\rangle=D_1|\Lambda_{as}\rangle \,,
\\
\label{mingtrsym}
&&
\delta_{sym}
|\Phi\rangle=(\frac{1}{2}D_1N_{21}-D_2)|\Lambda_{sym}\rangle\,.
\end{eqnarray}

\newcommand{\un}{\underline{n}}
\newcommand{\um}{\underline{m}}

\subsection{Gauge Symmetries in $AdS_d$}
\label{Gauge Symmetries in $AdS_d$}

Let us now analyze the situation in the AdS case.
The Lorentz covariant derivatives (\ref{lorspiope})
are no longer commuting but satisfy the commutation relationships
\be
\label{ads}
[ D_{\um} , D_{\un} ] =  -\lambda^2 M_{\um \un}\,,
\ee
\be
\label{tor}
D_{\um} \ga_{\un A}- D_{\un} \ga_{\um A}=
D_{\um} \bar{\ga}_{\un A}- D_{\un} \bar{\ga}_{\um A}=0
\ee
with the convention
\be
\ga_{\un A}= e_{\un}{}^n \ga_{n A}\,,\qquad
\bar{\ga}_{\un A}= e_{\un}{}^n \bar{\ga}_{n A}\,.
\ee
The  condition (\ref{tor}) is just the standard zero torsion condition
\be
D_{\um} e_{\un}{}^a- D_{\un} e_{\um }{}^a = 0
\ee
while (\ref{ads}) is the equation of the AdS space.
The covariant D'Alembertian  is
\begin{equation}\label{dalope}
{\cal D}^2\equiv
D_m^2+\omega_m{}^{mn}D_n\,,
\end{equation}
where the second term  accounts for $D_m$ being
rotated as a tangent vector. With these conventions the
covariant derivatives $D_A$ and $\bar{D}_B$ satisfy a number of
useful relationships summarized in the Appendix.

 As in the flat case we will analyze gauge symmetries with totally
symmetric and totally antisymmetric gauge parameters
 $\Lambda_{sym}^{mn}$ and $\Lambda_{as}^{mn}$.
It is sometimes convenient to combine them into  a
gauge parameter $\Lambda_{m,n}$ having no
definite symmetry properties
$$
|\Lambda\rangle=\Lambda_{m,n}\alpha_1^m\alpha_2^n|0\rangle\,.
$$
The symmetric and antisymmetric parts can be singled out as
\begin{equation}
\label{lsym}
|\Lambda_{sym}\rangle \equiv |S\rangle = N_{12}|\Lambda \rangle\,,
\end{equation}
\begin{equation}
\label{lasym}
|\Lambda_{as}\rangle\equiv (1-\frac{1}{2}N_{12}N_{21})|\Lambda\rangle\,.
\end{equation}

The gauge transformation $\delta|\Phi\rangle$ which
respects the constraints (\ref{n11})-(\ref{n12}) is
\begin{equation}\label{gtr0}
\delta_\Lambda |\Phi \rangle =
D_1 |\Lambda \rangle - D_2 |\Lambda_{sym}\rangle\,.
\end{equation}
It can equivalently be written as a combination of
the gauge transformations with symmetric and antisymmetric
gauge parameters
\begin{equation}\label{gtr1}
\delta_\Lambda|\Phi\rangle=\delta_{as}|\Phi\rangle+\delta_{sym}|\Phi\rangle
\end{equation}
with the gauge transformations of the form (\ref{mingtras}) and
(\ref{mingtrsym}) but now with the derivatives $D_1$ and $D_2$ as in
$AdS_d$.

Next we analyze whether there exists a Lagrangian that
generalizes (\ref{minlagten}) (equivalently,
(\ref{minlag})) to $AdS_d$.
The most general deformation of (\ref{minlag})
to the AdS case without higher derivatives is of the form
\begin{eqnarray}\label{init}
L^{\Phi\Phi}
&=&
\frac{1}{2}\langle\Phi|
{\cal D}^2 -f\lambda^2
-D_1 \bar{D}_1 -D_2 \bar{D}_2
+\frac{3}{4} P_{11}(-{\cal D}^2 + g \lambda^2  )\bar{P}_{11}
\nonumber\\
&+&\frac{3}{4} (P_{11}\bar{D}_1^2 + D_1^2\bar{P}_{11}
+P_{11}D_2\bar{D}_2 \bar{P}_{11})|\Phi \rangle\,,
\end{eqnarray}
where $f$ and $g$ are arbitrary parameters.
A straightforward but rather tedious computation with the use of
the identities collected in the Appendix leads to the following
result
\bee
\label{varl}
\delta S^{\Phi\Phi} &=& -\lambda^2\int_{AdS_d} e \Big [  \langle\Phi|(f+3) D_1
+\frac{3}{2}(d-5-g) P_{11} \bar{D}_1 |\Lambda \rangle\nonumber\\
&+&\langle\Phi| (6-3d-f)D_2 +3(d-3) P_{21} \bar{D}_1
+\frac{3}{2}(1-d-g) P_{21}D_1 \bar{P}_{11} ) |S\rangle \Big ]\,.
\eee
{}From this expression it is clear that the freedom in the
parameters $f$ and $g$ is not enough to warrant an action invariant
under both types of symmetries. The best one can do is to find
a Lagrangian invariant either with respect to the gauge symmetry with the
parameter
$|\Lambda_{as}\rangle$ or the one with
$|\Lambda_{sym}\rangle$.
Note that at the level of equations of
motion in the Lorentz gauge an analogous phenomenon was observed for
redundant gauge symmetries in \cite{met95}. Since, according to the
general analysis of unitary representations of $AdS_d$ in \cite{met95},
the case with gauge invariance with respect to
$|\Lambda_{sym}\rangle$ does not lead to unitary dynamics,
we focus on the Lagrangian possessing the
$|\Lambda_{as}\rangle$ invariance.  {}From (\ref{varl}) it is
obvious that this is achieved by setting
\be
f=-3 \,,\qquad g= d-5\,,
\ee
since the antisymmetric part of the gauge parameter enters only via
the first term. Thus, we set

\begin{eqnarray}\label{lagpp}
L^{\Phi\Phi}
&=&
\frac{1}{2}\langle\Phi|
{\cal D}^2 +3\lambda^2
-D_1 \bar{D}_1 -D_2 \bar{D}_2
+\frac{3}{4} P_{11}(-{\cal D}^2 +\lambda^2 (d-5) )\bar{P}_{11}
\nonumber\\
&+&\frac{3}{4} (P_{11}\bar{D}_1^2 + D_1^2\bar{P}_{11}
+P_{11}D_2\bar{D}_2 \bar{P}_{11})|\Phi \rangle\,.
\end{eqnarray}

Because one of the gauge symmetries is lost, the Lagrangian (\ref{lagpp})
describes more degrees of freedom than the original flat-space
Lagrangian we started with. This is in agreement
with the general conclusion of Sect.\ref{Flat Space Pattern of $AdS$
Massless fields} that physical
d.o.f.  of massless AdS fields may not be described by an
irreducible representation of $o(d-2)$. The conjecture of Sect.\ref{Flat
Space Pattern of $AdS$
Massless fields} suggests that the flat space and the AdS dynamics can match
only once one starts with
specific (reducible) collections of fields in flat space. {}From the general
analysis of  Sect.\ref{Flat Space Pattern of $AdS$ Massless fields}
it follows that, in order to make the AdS deformation  consistent for the case
under consideration,  one has to add a massless spin two
field analogous to a graviton field.

Let us therefore introduce the field $\chi^{mn}$
symmetric in indices $m,n$, described by the Fock vector
$$
|\chi\rangle\equiv \chi_{mn}\alpha_1^m\alpha_1^n|0\rangle\,.
$$
Since this field should describe a massless
spin 2 field in the flat limit, it has its own gauge symmetry with the gauge
parameter
\be
|\xi\rangle\equiv \xi_m \alpha_1^m|0\rangle \,.
\ee
The idea is that starting from the sum of the free Lagrangians
$L^{\Phi\Phi}+L^{\chi\chi}$ one should add
cross terms $L^{\Phi\chi}$
which (i) reestablish all (appropriately deformed by $\lambda$-dependent
terms) gauge symmetries with the parameters
$|\Lambda \rangle$ and $|\xi\rangle$ and (ii)
tend to zero in the flat limit.
It turns out that this is indeed possible.
The final result is that the action
\begin{equation}\label{comlag}
S=\int_{AdS_d} e [L^{\Phi\Phi}+L^{\Phi\chi}+L^{\chi\chi} ]\,,
\end{equation}
where
\begin{eqnarray}
L^{\Phi\chi}
&=&
\frac{3}{2}(d-3)\lambda\langle \Phi |-2 D_2 +2 P_{12}\bar{D}_1
+P_{11}D_2 \bar{P}_{11} |\chi\rangle\,,
\\
L^{\chi\chi}
&=&\frac{3}{2}(d-3) \langle\chi|
{\cal D}^2 +d \lambda^2 - D_1\bar{D}_1
-\frac{1}{2} P_{11} ({\cal D}^2 +\lambda^2)\bar{P}_{11}
\nonumber\\
&+&
\frac{1}{2}(P_{11} \bar{D}_1^2 +D_1^2\bar{P}_{11} )|\chi\rangle\,
\end{eqnarray}
is invariant under the gauge transformations of the form
\begin{eqnarray}
&&
\delta |\Phi \rangle =
D_1 |\Lambda \rangle -D_2 |S\rangle
+\lambda (P_{12} -P_{11} N_{21})|\xi \rangle\,,
\\
&&
\label{dch}
\delta |\chi \rangle =D_1 |\xi \rangle +\lambda |S\rangle\,.
\end{eqnarray}

Note that the Lagrangian $L^{\Phi \chi}$ is proportional to $\lambda$
and tends to zero in the flat limit. Therefore, as expected,
the action reduces in the flat limit
to the sum of two actions for the irreducible fields. In the AdS
case, however, the cross term   $L^{\Phi \chi}$ becomes nontrivial
so that  the system does not decompose into a sum of
elementary subsystems.
Another comment is that according to (\ref{dch}) the field
$|\chi \rangle$ becomes a Stueckelberg field that can be gauged
away for $\lambda \neq 0$. The resulting gauge fixed action is
nothing but the
action (\ref{lagpp}) invariant under the gauge symmetry
with antisymmetric gauge parameter. Therefore it describes properly
the irreducible AdS representation. The gauge fixing
$|\chi \rangle=0$ is impossible however for $\lambda = 0$. This is
why the naive
flat limit of the action (\ref{lagpp}) describes not
two fields  but only one in agreement with
\cite{AKO}. This phenomenon can be interpreted as some sort
of nonanalyticity of the flat limit exhibited already at the
free field level.

We expect that one can analogously find a deformation
of the flat space Lagrangian (\ref{minlag})
to  $AdS_d$ by adding
an antisymmetric second rank gauge tensor. This would correspond
to keeping the symmetry with symmetric parameters. However, because
the corresponding representation of the AdS algebra is not unitary,
the resulting gauge invariant Lagrangian is expected to have a wrong
 relative  sign  of the kinetic terms
of the elementary flat-space Lagrangian (i.e., incompatible with
unitarity). A similar phenomenon is expected to be true for more complicated
Young diagrams: only the sets of fields predicted in Sect.\ref{Flat Space
Pattern of $AdS$
Massless fields} will have all signs of kinetic terms of the flat-space
Lagrangians correct.

The equations of motion for the fields $|\Phi\rangle$ and $|\chi\rangle$
that follow  from the Lagrangian (\ref{comlag}) can be reduced to the
form
\begin{eqnarray}
\label{eqmotphi}
\Bigl({\cal D}^2
-D_1\bar{D}_1-D_2\bar{D}_2+\frac{1}{2}D_1^2\bar{P}_{11}
+D_2D_1\bar{P}_{12}
-\lambda^2P_{11}\bar{P}_{11}
- 2\lambda^2 P_{12}\bar{P}_{12} + 3\lambda^2
\Bigr)|\Phi\rangle
\nonumber\\
+\lambda\Bigl(
(d-3)(D_1N_{21}-2D_2)
-P_{12}\bar{D}_1+P_{11}N_{21}\bar{D}_1
+(P_{12}D_1-P_{11}D_2)\bar{P}_{11}\Bigr)|\chi\rangle=0\,,
\end{eqnarray}

\begin{equation}\label{eqmotchi}
({\cal D}^2-D_1\bar{D}_1+\frac{1}{2}D_1^2\bar{P}_{11}
-\frac{1}{2}P_{11}\bar{P}_{11} +  d\lambda^2)|\chi\rangle
+\lambda(\bar{D}_2-D_1\bar{P}_{12})|\Phi\rangle=0\,.
\end{equation}
By imposing the Lorentz gauge  $\bar{D}_i|\Phi\rangle=0$,
the tracelessness condition $\bar{P}_{ij}|\Phi\rangle=0$ and the
condition $|\chi\rangle=0$, we are left with
\begin{equation}\label{eqmot1}
({\cal D}^2+3\lambda^2)|\Phi\rangle=0\,.
\end{equation}
There is a leftover symmetry with the parameter
$|S\rangle$ satisfying certain differential conditions.
Taking into account that one can identify the Stueckelberg
field $|\chi\rangle$ with $|S\rangle$ one can derive these
conditions from the equations of motion for $|\chi\rangle$
in the  gauge $\bar{D}_i|\chi\rangle=0$, $\bar{P}_{ij}|\chi\rangle=0$
\begin{equation}\label{eqmot2}
({\cal D}^2+d\lambda^2)|S\rangle=0\,.
\end{equation}

Let us compare  these results with the
equations for the gauge field corresponding  to  the $AdS_d$  massless
representation $D(E_0,{\bf s})$
\begin{equation}\label{gt1}
({\cal D}^2- E_0(E_0+\lambda-d\lambda)+\lambda^2\sum_{A=1,2}
s_A)|\Phi\rangle=0\,, \end{equation}
and the conditions on the leftover gauge
parameter $|S\rangle$
\begin{equation}\label{gt2}
\Big ({\cal D}^2
-\lambda^2(s_2-2)(s_2-3+d)+\lambda^2\sum_{A=1,2}s_A
-\lambda^2 \Big ) |S\rangle
=0\,
\end{equation}
found in \cite{met95}.
For the case under consideration the $E_0$ and ${\bf s}$
are
$$
E_0=\lambda(d-1)\,,
\qquad
{\bs}=(2,1,0,\ldots,0)\,.
$$
Plugging these values into (\ref{gt1}) (\ref{gt2}) we indeed
arrive at the equations (\ref{eqmot1}) and (\ref{eqmot2}).

\section{Conclusions}

We have shown that generic
irreducible massless (gauge) fields in $AdS_d$
in the flat limit decompose into nontrivial sets of irreducible
flat space massless fields. These sets are, however, smaller
than the result of a dimensional reduction to one less dimension of a
corresponding massive field. In that sense $AdS_d$ massless fields
are ``less massless" than flat space massless fields. We made a
conjecture on the pattern of the flat-space reduction of a generic
$AdS_d$ massless field. From this conjecture it follows that
there is a unique nontrivial situation when a flat space spin
two massless field appears as a result of a nontrivial reduction
of $AdS_d$ massless field with mixed symmetry properties. This
example has been considered in detail. It is tempting to speculate that
there may exist some new version of gravity associated with this type
of field.

On the other hand we have
argued that totally antisymmetric tensors can never
result from the  flat limit decomposition of other types
of $AdS_d$ unitary
representations. In other words, the space of
differential
forms is closed with respect to the flat space limit decomposition.

An interesting problem for the future is to generalize these results
to the supersymmetric cases to analyze generic $AdS$ supermultiplets
in higher dimensions and, in particular, in $AdS_{11}$. Another problem is to
consider the multiplets occurring in \cite{pr} to see how they group
themselves in the case
of $AdS$.


\section*{Acknowledgments}

The work of R. Metsaev and M. Vasiliev is supported in part by
INTAS, Grant No.96-0538 and Grant No.99-0590 and
by the RFBR Grant No.99-02-16207. 
R. Metsaev is also supported by DOE/ER/01545-787.
M. Vasiliev was also supported by NFR
F-FU 08115-347. The work of
L. Brink is supported by NFR F 650-19981268/2000. R. Metsaev and M. Vasiliev
would like to thank for
hospitality  at Chalmers University. M. Vasiliev would also 
like to express his gratitude
to Prof. H.Nicolai for the hospitality at the
MPI f\"ur Gravitationsphysik, Albert Einstein Institute where
some part of this work was done. The authors are grateful to O.Gelfond
for the help in \LaTeX  drawing of Young diagrams.

\section*{Appendix. Algebra of commutators}

In this appendix we collect some formulas that are used
in the computations of the section \ref{Gauge Symmetries in $AdS_d$}.

\begin{equation}\label{bdidj}
[\bar{D}_A,D_B]=\delta_{AB}({\cal D}^2
+\lambda^2\sum_C N_{CC})+\lambda^2(1-d)N_{BA}
+\lambda^2\sum_C(P_{BC}\bar{P}_{AC} - N_{CA} N_{BC})\,,
\end{equation}
$$
[D_A,D_B]=\lambda^2\sum_C(P_{BC}N_{AC}-P_{AC}N_{BC})\,,
$$
$$
[\bar{D}_A,\bar{D}_B]=\lambda^2\sum_C(N_{CB}\bar{P}_{AC}
-N_{CA}\bar{P}_{BC})\,,
$$
$$
[{\cal D}^2,D_A]=\lambda^2(1-d)D_A
+2\lambda^2\sum_C (P_{AC}\bar{D}_C-D_C N_{AC})\,,
$$
$$
[{\cal D}^2,\bar{D}_A]=\lambda^2(d-1)\bar{D}_A
+2\lambda^2\sum_C (N_{CA}\bar{D}_C-D_C\bar{P}_{AC})\,,
$$
where ${\cal D}^2$ in (\ref{bdidj}) is a covariant D'Alembertian operator
(\ref{dalope}). These formulas can be derived by straightforward but
sometimes lengthy calculation.
The derivation of the following  relationships:
$$
[N_{AB},N_{CE}]=\delta_{BC}N_{AE}-\delta_{AE}N_{CB}\,,
$$

$$
[\bar{D}_A,N_{BC}]=\delta_{AB}\bar{D}_C\,,
\qquad
[N_{AB},D_C]=\delta_{BC}D_A\,,
$$

$$
[\bar{D}_A,P_{BC}]=\delta_{AB}D_C+\delta_{AC}D_B\,,
$$
$$
[\bar{P}_{AB},D_C]=\delta_{BC}\bar{D}_A+\delta_{AC}\bar{D}_B\,,
$$

$$
[\bar{P}_{AB},P_{CE}]
=\delta_{BC}N_{EA}+\delta_{BE}N_{CA}
+\delta_{AC}N_{EB}+\delta_{AE}N_{CB}
+d(\delta_{BC}\delta_{AE}+\delta_{BE}\delta_{AC})\,.
$$
is elementary.


\begin{thebibliography}{100}

\bibitem{Fr}
C. Fronsdal,
{\it Phys. Rev.\/} {\bf D18} (1978) 3624;
{\bf D20} (1979) 848;

J. Fang and C. Fronsdal,
{\it Phys. Rev.\/} {\bf D18} (1978) 3630;
{\bf D22} (1980) 1361

\bibitem{vas}
M.A. Vasiliev,
{\it Fortschr. Phys.\/} {\bf 35} (1987) 741; ``Higher Spin Gauge Theories:
Star-Product and AdS Space",
Contributed article to Golfand's Memorial Volume, M. Shifman ed., World
Scientific. hep-th/9910096

\bibitem{met08}
R.R. Metsaev,
``IIB supergravity and various aspects of
light-cone formalism in AdS space-time'',
Based on talk given at International Workshop
"Supersymmetries and Quantum Symmetries",
Dubna, Russia, July 27-31, 1999;
hep-th/0002008
\bibitem{FV}

E.S.Fradkin and M.A. Vasiliev,
\AP{177}{1987}{63}

\bibitem{van}
M.A. Vasiliev,
\AP{190}{1989}{59}

\bibitem{FV1}
E.S. Fradkin and M.A. Vasiliev,
\PLB{189}{1987}{89};
\NPB{291}{1987}{141}


\bibitem{mal}
J. Maldacena,
\ATMP{2}{1998}{231},
hep-th/9711200;
S.S. Gubser, I.R. Klebanov, A.M. Polyakov,
\PLB{428}{1998}105,
hep-th/9802109;
E. Witten,
\ATMP{2}{1998}253,
hep-th/9802150

\bibitem{fer}
M.A. Vasiliev,
\NPB{301}{1988}{26}

\bibitem{vas2}
M.A. Vasiliev,
\PLB{257}{1991}{111}

\bibitem{op}
M.A. Vasiliev,
{\it Fortschr. Phys.} {\bf 36} (1988) 33

\bibitem{vas1}
M.A. Vasiliev,
\PLB{243}{1990}{378};
\CQG{8}{1991}{1387};
\PLB{285}{1992}{225}

\bibitem{pr}
T. Pengpan, P. Ramond,
\PR{315}{1999}{137};
hep-th/9808190


\bibitem{br}
L. Brink, P. Ramond,
``Dirac Equations,
Light Cone Supersymmetry, and Superconformal Algebras''
Contributed article to Golfand's Memorial Volume, M. Shifman ed., World
Scientific.
hep-th/9908208


\bibitem{Nic}
H. Nicolai,
``Representations of supersymmetry in anti-de
Sitter space'', in: Supersymmetry and supergravity, '84,
Ed. B. de Wit, P. Fayet and P. van Nieuwenhuizen
(World Scientific, Singapore, 1984)

\bibitem{dewit}
B. de Wit, I. Herger,
``Anti-de Sitter Supersymmetry'',
Lectures given at 35th Winter School of Theoretical Physics: From
Cosmology to Quantum Gravity, Polanica, Poland, 2-12 Feb 1999;
hep-th/9908005

\bibitem{F1}
C. Fronsdal,
\PRD{12}{1975}3819

\bibitem{metlc}
R.R. Metsaev,
\NPB{563}{1999}{295};
hep-th/9906217

\bibitem{evans}
N.T. Evans,
\JMP{8}{1967}{170}

\bibitem{dirac}
P.A.M. Dirac
\JMP{4}{1963}901

\bibitem{murat}
M. Gunaydin,
``Singleton and doubleton supermultiplets of space-time supergroups
and infinite
spin superalgebras'',
Invited talk given at Trieste Conf. on Supermembranes and Physics in
(2+1)-Dimensions, Trieste, Italy, Jul 17-21, 1989

\bibitem{gunmin}
M. Gunaydin, D. Minic,
\NPB{523}{1998}{145};
hep-th/9802047

\bibitem{met95}
R.R. Metsaev,
\PLB{354}{1995}{78};
R.R. Metsaev,
``Arbitrary spin massless bosonic
fields in d-dimensional anti-de Sitter space'',
Talk given at Dubna International Seminar
`Supersymmetries and Quantum Symmetries'  dedicated to
the memory of Victor I. Ogievetsky, Dubna, 22-26 July, 1997;
hep-th/9810231


\bibitem{mack}
G. Mack,
\CMP{55}{1977}{1}


\bibitem{LV}
V.E.Lopatin and M.A.Vasiliev,
\MPLA{3}{1988}{257}

\bibitem{labas}J.M.F.Labastida, \NPB{322}{1989}{185}

\bibitem{barut}
A.O. Barut and R. Raczka,
Theory of Group Representations and Applications
(PWN-Polish Scientific Publishers, Warszawa 1977)


\bibitem{AKO}
C.S.Aulakh, I.G.Koh, and S.Ouvry,
\PLB{173}{1989}{284}

\end{thebibliography}
\end{document}